\documentclass[preprint,prb,showpacs,superscriptaddress]{revtex4-1}
\usepackage{amsmath}
\usepackage{graphicx}
\usepackage{amssymb}
\usepackage{times}
\usepackage{color}
\usepackage{multirow}
\definecolor{lr}{rgb}{1.0,0.3,0.3}
\definecolor{dg}{rgb}{0.0,0.5,0.0}

\usepackage{tabularx} 

\usepackage{comment}

\graphicspath{./}


\begin{document}

\title{Photoluminescence at the ground state level anticrossing of the nitrogen-vacancy center in diamond}

\author{Viktor Iv\'ady}
\email{ivady.viktor@wigner.hu}
\affiliation{Wigner Research Centre for Physics,
  PO Box 49, H-1525, Budapest, Hungary}
\affiliation{Department of Physics, Chemistry and Biology, Link\"oping
  University, SE-581 83 Link\"oping, Sweden}

\author{Huijie Zheng}
\email{zheng@uni-mainz.de}
\affiliation{Johannes Gutenberg-Universit{\"a}t Mainz, 55128 Mainz, Germany}
 \affiliation{Helmholtz-Institut, GSI Helmholtzzentrum f{\"u}r Schwerionenforschung, 55128 Mainz, Germany}

\author{Arne Wickenbrock}
\affiliation{Johannes Gutenberg-Universit{\"a}t Mainz, 55128 Mainz, Germany}
 \affiliation{Helmholtz-Institut, GSI Helmholtzzentrum f{\"u}r Schwerionenforschung, 55128 Mainz, Germany}

\author{Lykourgos Bougas}
\affiliation{Johannes Gutenberg-Universit{\"a}t Mainz, 55128 Mainz, Germany}

\author{Georgios Chatzidrosos}
\affiliation{Johannes Gutenberg-Universit{\"a}t Mainz, 55128 Mainz, Germany}

\author{Kazuo Nakamura} 
\affiliation{Leading-Edge Energy System Research Institute, Fundamental Technology Dept., Tokyo Gas Co., Ltd., Yokohama, 230-0045 Japan}

\author{Hitoshi Sumiya} 
\affiliation{Advanced Materials Laboratory, Sumitomo Electric Industries, Ltd., Itami, 664-0016 Japan}

\author{Takeshi Ohshima} 
\affiliation{Takasaki Advanced Radiation Research Institute, National Institutes for Quantum and Radiological Science and Technology, Takasaki, 370-1292, Japan }

\author{Junichi Isoya}
\affiliation{Faculty of Pure and Applied Sciences, University of Tsukuba, Tsukuba, 305-8573 Japan}

\author{Dmitry Budker}
\affiliation{Johannes Gutenberg-Universit{\"a}t Mainz, 55128 Mainz, Germany}
 \affiliation{Helmholtz-Institut, GSI Helmholtzzentrum f{\"u}r Schwerionenforschung, 55128 Mainz, Germany}
\affiliation{Department of Physics, University of California, Berkeley, CA 94720-7300, USA }

\author{Igor A. Abrikosov} 
\affiliation{Department of Physics, Chemistry and Biology, Link\"oping
  University, SE-581 83 Link\"oping, Sweden}

\author{Adam Gali}
\affiliation{Wigner Research Centre for Physics,
  PO Box 49, H-1525, Budapest, Hungary}
\affiliation{Department of Atomic Physics, Budapest University of Technology and Economics, Budafoki \'{u}t 8., H-1111 Budapest, Hungary}

\date{\today}

\begin{abstract}
The nitrogen-vacancy center (NV center)  in diamond at magnetic fields corresponding to the ground state level anticrossing (GSLAC) region gives rise to rich photoluminescence (PL) signals due to the vanishing  energy gap between the electron spin states, which enables to have an effect on the NV center's luminescence for a broad variety of environmental couplings. In this article we report on the GSLAC photoluminescence signature of NV ensembles in different spin environments at various external fields. We investigate the effects of transverse electric and magnetic fields, P1 centers, NV centers, and the $^{13}$C nuclear spins, each of which gives rise to a unique PL signature at the GSLAC. The comprehensive analysis of the couplings and related optical signal at the GSLAC provides a solid ground for advancing various microwave-free applications at the GSLAC, including but not limited to magnetometry, spectroscopy, dynamic nuclear polarization (DNP), and nuclear magnetic resonance (NMR) detection. We demonstrate that not only the most abundant $^{14}$NV center but the $^{15}$NV can also be utilized in such applications and that nuclear spins coupled to P1 centers can be polarized directly by the NV center at the GSLAC, through a giant effective nuclear $g$-factor arising from the NV center-P1 center-nuclear spin coupling. We report on new alternative for measuring defect concentration in the vicinity of NV centers and on the optical signatures of interacting, mutually aligned NV centers.

\end{abstract}
\maketitle

%%%%%%%%%%%%%%%%%%%%%%%%%%%%%%%%%%%%%%%%%%%%%%%%%%%%%%%%%%%%%%%%%%%%%%%%%%%%%%%
%%%%%%%%%%%%%%%%%%%%%%%%%%%%%%%%%%%%%%%%%%%%%%%%%%%%%%%%%%%%%%%%%%%%%%%%%%%%%%%

\section{Introduction}

Over the last decade, the NV center in diamond\citep{duPreez:1965,Wrachtrup:JPCM2006,Maze2011,DohertyNVreview} has demonstrated considerable potential in spectroscopy and sensing applications.\citep{Maze:Nature2008,Balasubramanian:NatMat2009,Dolde2011,Kucsko2013,Teissier2014} The NV center exhibits a level anticrossing in the electronic ground state (GSLAC)  at  magnetic field $B_z = \pm 102.4$~mT,  which has been recently exploited in microwave-free applications, ranging from magnetometry\citep{BroadwayPhysRevApplied2016,WickenbrockAPL2016,Huijie2020} through nuclear-magnetic-resonance spectroscopy\citep{EpsteinNatPhys2005,Wang2014,BroadwayPhysRevApplied2016,WoodNatComm2017} to optical hyperpolarization\citep{HaiJing2013,IvadyPRL2016,BroadwayNatComm2018}.  These applications are of special interest in biology and medicine, where high-power microwave driving is undesirable. 

The physics of isolated NV centers at the GSLAC is well-understood,\citep{BroadwayPhysRevApplied2016,IvadyPRL2016} however, the effects of environmental perturbations are not comprehensively described. Due to the presence of $^{14}$N ($^{15}$N) nuclear spin, six (four) mixed electron-nuclear spin states either cross or exhibit  an avoided crossing. External perturbations and interaction with the local nuclear and electron spin environment may give rise to additional spin-relaxation mechanisms at specific magnetic fields corresponding to the crossings of the spin states. Through the spin-dependent PL of the NV center, these processes may give rise to various PL signals at the GSLAC \citep{Armstrong2010,Anishchik2017,Anishchik2019,Huijie2020}.  Besides the optical signal, optically detected magnetic resonance (ODMR) signal of a NV ensemble has been recently recorded.\citep{Auzinsh19}  Recently, the ground state level anticrossing at zero magnetic field and related phenomena have attracted considerable  attention.\citep{Hannah2016,Dmitriev2019} As increasing number of  applications rely on LAC signals of single or ensemble NV systems, quantitative description of the most relevant environmental  couplings is essential for further development and engineering of these applications.

Furthermore, interaction between NV centers and $^{13}$C nuclear spins at the GSLAC can potentially be utilized in dynamic nuclear polarization (DNP)\citep{Jacques2009,IvadyDNP2015,Sosnovsky2019} applications. DNP can give rise to a hyperpolarized diamond sample with a potential to transfer spin polarization to adjacent nuclear spins for the improvement of traditional nuclear magnetic resonance methods\citep{HaiJing2013,PlenioNanoLett2018,BroadwayNatComm2018,AjoySciAdv2018,SchwartzSciAdv2018,ShagievaNanoLett2018,ZangaraNanoLett2019}.  It is therefore of fundamental importance to gain detailed insight into the NV-$^{13}$C spin dynamics at the GSLAC.

In this article we aim at establishing a guideline for developing and advancing applications at the GSLAC of the NV center in diamond by collecting and describing the most relevant interactions that may either limit existing applications or give rise to new ones. Indeed, by identifying the PL signals of different environmental couplings we reveal important interactions that enables new spectroscopy, magnetometry and dynamic nuclear polarization applications

The rest of the paper is organized as follows: in section~\ref{sec:bg} we provide a brief overview of the established physics of the NV center. In section~\ref{sec:meth}, we describe our experimental setup and samples and the details of the theoretical simulations. Section~\ref{sec:res} describes our results in four sections considering interactions of NV centers with external fields, $^{13}$C nuclear spins, P1 centers, and other NV centers at the GSALC. In section~\ref{sec:disc}, we discuss implications of our results. Finally, we summarize the findings in section~\ref{sec:sum}.

\section{Background}
\label{sec:bg}

The NV center in diamond gives rise to a coupled hybrid register that consists of a spin-1 electron spin and either a spin-1 $^{14}$N or spin-1/2 $^{15}$N. Hereinafter, we refer to the former as $^{14}$NV center and to the latter, less abundant configuration, as $^{15}$NV center. 

The spin Hamiltonian of the $^{14}$NV center can be written as
\begin{equation} \label{eq:H0_14NV}
H_{^{14}\text{N}} =  D \left( S_z^2 - \frac{ 2 }{3} \right) + g_{\text{e}} \beta S_z B_z + S \mathcal{A}_{^{14}\text{N}} I_{^{14}\text{N}} + Q \left( I_{^{14}\text{N}, z}^2 - \frac{ 2 }{3} \right) - g_{^{14}\text{N}} \beta_{\text{N}} I_{^{14}\text{N}, z} B_z \text{,}
\end{equation}
where terms on the right-hand-side describe zero-field splitting, Zeeman, hyperfine, nuclear quadrupole, and nuclear Zeeman interaction, respectively, $S$ and $I_{^{14}\text{N}}$ are the electron and nuclear spin operator vectors, and $S_z$ and $I_{^{14}\text{N}, z}$ are the electron and nuclear spin $z$ operators, where the quantization axis $z$ is parallel to the N-V axis. $g_{\text{e}}$ and $g_{^{14}\text{N}}$ are the electron and $^{14}$N nuclear $g$-factors, $\beta$ and $\beta_{\text{N}}$ are the Bohr and nuclear magnetons, respectively, $D = 2868.91$~MHz is the zero-field splitting, $Q = -5.01$~MHz\citep{Felton:PRB2009} is the nuclear quadrupole splitting, and  $\mathcal{A}^{^{14}\text{N}}$ is the hyperfine tensor of the $^{14}$N nuclear spin that can be expressed by its diagonal elements $ A_{xx} = A_{yy} = A_{\perp} = -2.70$~MHz and $A_{zz} = A_{\parallel} = -2.14$~MHz\citep{Felton:PRB2009}. 

The spin Hamiltonian of the $^{15}$NV center can be written as
\begin{equation} \label{eq:H0_15NV}
H_{^{15}\text{N}} =  D \left( S_z^2 - \frac{ 2 }{3} \right) + g_{\text{e}} \beta S_z B_z + S \mathcal{A}_{^{15}\text{N}} I_{^{15}\text{N}} - g_{^{15}\text{N}} \beta_{\text{N}} I_{^{15}\text{N}, z} B_z \text{,}
\end{equation}
where $g_{^{15}\text{N}}$ is the nuclear $g$-factor of $^{15}$N nucleus, and  $\mathcal{A}_{^{15}\text{N}}$ is the hyperfine tensor of the $^{15}$N nuclear spin that can be expressed by its non-zero diagonal elements $ A_{\perp} = +3.65$~MHz and $ A_{\parallel} = +3.03$~MHz\citep{Felton:PRB2009}. 

Diamond contains 1.07\% spin-1/2 $^{13}$C isotope  in natural abundance  that can effectively interact with the NV electron spin at the GSLAC through the hyperfine interaction. The Hamiltonian of a $^{13}$C nuclear spin  coupled to a NV center can be written as
\begin{equation}
\hat{H}_{^{13}\text{C}} =  g_{^{13}\text{C}} \mu_{N} B \hat{I}_{^{13}\text{C},z} + \hat{S}A_{^{13}\text{C}}\hat{I}_{^{13}\text{C}} \text{,}
\end{equation}
where $\hat{I}_{^{13}\text{C}}$ is the nuclear spin operator vector, $g_{^{13}\text{C}} $ is the nuclear $g$-factor of $^{13}$C nucleus, and $A_{^{13}\text{C}}$ is the hyperfine tensor that consists of two terms, the isotropic Fermi contact term and the anisotropic dipolar interaction term,
\begin{equation}
A_{^{13}\text{C}} = A^{\text{Fc}}_{^{13}\text{C}} + A^{\text{d}}_{^{13}\text{C}}  \text{.}
\end{equation}
Due to the typically low symmetry of the NV-$^{13}$C coupling, all the six independent elements of the hyperfine tensor  can be non-zero in the coordinate system of the NV center. These components can be expressed by the diagonal hyperfine tensor elements, $A_{xx} \approx A_{yy} = A_{\perp}$ and $A_{zz} = A_{\parallel}$, as well as angle $\theta$ of the principal hyperfine axis $\mathbf{e}_z$ and the symmetry axis of the NV center. The hyperfine Hamiltonian, expressed in the basis of $\left | m_{\text{S}}, m_{ ^{13} {\text{C}}}   \right\rangle = \left \{  \left | 0, \uparrow   \right\rangle,  \left | 0, \downarrow   \right\rangle, \left | -1, \uparrow   \right\rangle, \left | -1, \downarrow   \right\rangle   \right \} $, can be written as
\begin{equation} \label{eq:hypop}
\hat{H}_{^{13}\text{C}} = \hat{S} A_{^{13}\text{C}} \hat{I}_{^{13}\text{C}} = \frac{1}{2}
\begin{pmatrix}
 0 & 0 & \frac{1}{\sqrt{2}} b & \frac{1}{\sqrt{2}} c_{-} \\
 0 & 0 & \frac{1}{\sqrt{2}}  c_{+} & -\frac{1}{\sqrt{2}} b \\
\frac{1}{\sqrt{2}} b & \frac{1}{\sqrt{2}}  c_{+} & -A_z & -b \\
\frac{1}{\sqrt{2}} c_{-} & -\frac{1}{\sqrt{2}} b & -b & A_z \\
 \end{pmatrix}
 \text{,}
\end{equation}
 where
 \begin{eqnarray} \label{eq:hyp_par1}
 A_z =  A_{\parallel} \cos^2 \theta  + A_{\perp} \sin^2 \theta \text{,}  \\
 \label{eq:hyp_par2}
 b = \left( A_{\parallel} - A_{\perp} \right) \cos \theta \sin \theta \text{,}   \\
 \label{eq:hyp_par3}
 c_{\pm} =  A_{\parallel} \sin^2 \theta   + A_{\perp}  \left( \cos^2 \theta  \pm 1  \right) \text{.}
\end{eqnarray}
Parameters $A_z$ and $b$ of the hyperfine Hamiltonian describe effective longitudinal and transverse magnetic fields due to the interaction with the $^{13}$C nuclear spin, respectively, while parameters $c_{+}$  and $c_{-}$ are responsible for mutual spin flip-flops of the $^{13}$C nuclear spin and the NV center. Note that there are non-zero matrix elements that correspond, for example, to $S_{+} I^{^{13}C}_+$ or $S_{+} I^{^{13}C}_z$ operator combinations. Appearance of such terms imply that $m_{\text{S}} $ and $ m_{ ^{13} {\text{C}}}$ are no longer good quantum numbers.

Besides nuclear spins, the NV center can interact with other spin defects, such as the spin-1/2 nitrogen substitution point defect (P1 center) and other NV centers. The spin Hamiltonian of the P1 center can be written as
\begin{equation}\label{eq:HP1}
H_{P1} =  g_{\text{e}} \beta S^{\text{P1}}_{z'} B_z + S^{\text{P1}} \mathcal{A}^{\text{P1}}_{z'} I^{\text{P1}} + Q^{\text{P1}}_{z'} \left( \left( I^{\text{P1}}_{z'} \right) ^2 - \frac{ 2 }{3} \right) - g_{\text{N}} \beta_{\text{N}} I^{\text{P1}}_{z'} B_z \text{,}
\end{equation}
where $\mathcal{A}^{\text{P1}}$ is the hyperfine interaction tensor that can be expressed by $A^{\text{P1}}_{\perp} = 81$~MHz and $A^{\text{P1}}_{\parallel} = 114$~MHz diagonal elements. For simplicity the quadrupole interaction strength is set to the value of the NV centers quadrupole splitting in this article, i.e.\ $ Q^{\text{P1}} = -5.01$~MHz, which is comparable with the measured  quadrupole splitting of $-3.974$~MHz of the P1 center\citep{Cook66}. Both the P1 center and other NV centers may exhibit a distinct local quantization axis depending on the C$_{3v}$ reconstruction and the N-V axis, respectively. We denote the symmetry and quantization axis of the P1 center in Eq.~(\ref{eq:HP1}) by ${z}'$.  The angle between ${z}'$ and the quantization axis of the central NV center can be either $0^{\circ}$ or $109.5^{\circ}$. The spin Hamiltonian of NV center that has ${z}'$ orientation can be obtained from Eq.~(\ref{eq:H0_14NV}) by a proper transformation of the coordinate system.

The interaction Hamiltonian between paramagnetic defects and the central NV center can be written as
\begin{equation} \label{eq:opJ}
H_{J} = S \mathcal{J} S^{\text{def}} \text{,}
\end{equation}
where $S^{\text{def}}$ is the spin operator vector of the spin defect and  $\mathcal{J}$ is the coupling tensor. Assuming point like electron spin densities, $\mathcal{J}$ can be approximated by the dipole-dipole coupling tensor.

\section{Methodology}
\label{sec:meth}

\subsection*{Theoretical approaches}

We employ two different theoretical approaches to study the GSLAC photoluminescence signal of NV ensembles interacting with external fields and environmental spins. For external fields, the density matrix $\varrho $ of a single NV center is propagated over a finite time interval according to the master equation of the closed system,
\begin{equation} \label{eq:master}
\dot{\varrho} = - \frac{i}{\hbar} \left[ H, \varrho \right] \text{,}
\end{equation}
where  $H$  is the ground-state spin Hamiltonian specified in section~\ref{sec:bg}. The starting density matrix $ \varrho_0 $ is set to describe 99.99\% polarization in the $\left| m_{\text{S}}, m_{^{14}\text{N}}\right\rangle =  \left| 0, +1 \right\rangle$ state of the electron and the $^{14}$N nuclear spins of the NV center. The PL intensity $\mathcal{I}$ is  approximated from the time averaged density matrix according to the formula of
\begin{equation} \label{eq:PLi}
\mathcal{I} \approx \left\langle p_0 \right\rangle + ( 1 - C ) \left\langle p_{\pm1} \right\rangle  \text{,}
\end{equation}
where the $ \left\langle p_0 \right\rangle $ and $ \left\langle p_{\pm1} \right\rangle $ are the time averaged probabilities of finding the electron spin in $m_{\text{S}} = 0$ and $m_{\text{S}} = \pm1$, respectively, and $C = 0.3$ is a reasonably experimentally attainable ODMR contrast. 

To study  the effects of environmental spins on the GSLAC photoluminescence signal, we apply a recently developed extended Lindblad formalism\citep{IvadyPRb2020}. In this approach spin-relaxation of a selected point defect surrounded by a bath of environmental spins can be simulated over either a fixed simulation time or cycles of ground state time evolution and optical excitation steps. The modeled systems consist of a central NV center, either $^{14}$NV or $^{15}$NV, and a bath of coupled environmental spins of the same kind. Different bath spins considered in our study are spin-1/2 $^{13}$C, spin-1/2 P1 center with a spin-1 $^{14}$N nuclear spin, and spin-1 $^{14}$NV and $^{15}$NV centers.  To create a realistic spin bath, spin defects are distributed randomly in the diamond lattice around the central NV center in a sphere.  To obtain ensemble averaged PL spectra, in all cases, we consider an ensemble of configurations, i.e.\  a set of random distributions of the spin defects. While each configuration describes a different local environment of the NV center, the ensembles describe a certain spin bath concentration on average. As a main approximation of the method, the many-spin system is divided into a cluster of subsystems. The number of spins included in each cluster determine the order of the cluster approximation. In the first-order cluster approximation no entanglement between the bath spins is taken into account. Higher order modeling allows inclusion of intra-spin bath entanglement. For further details on the methodology 
see Ref.~[\onlinecite{IvadyPRb2020}]. For simplicity, the mean field of the spin 
bath\citep{IvadyPRb2020} is neglected in this study.

In the case of a $^{13}$C spin bath, the nuclear spin-relaxation time is long compared to the inverse of the optical pump rate, which enables nuclear spin polarization to play considerable role in the GSLAC PL signal of NV centers. Therefore, to simulate the PL signal we simulated a sequence of optical excitation cycles. Each of them included two steps, 1) coherent time evolution in the ground state with a dwell time $t_{\text{GS}}$ set to 3~$\mu$s, and 2) spin selective optical excited process taken into account by a projection operator defined as
\begin{equation}
D =  \left( 1 - C \right) I + C p_{\pm} P_{\pm1 \rightarrow 0} + C p_{0} P_{ 0 \rightarrow 0 }\text{,}
\end{equation}
where $I$ is the identity operator, $P_{i \rightarrow f}$ is a projector operator from  $\left| m_S = i  \right\rangle$ state to  $\left| m_S = f  \right\rangle$ state of the NV spin, and $p_{s}$ is the probability of finding the system in state $\left| m_S = s \right\rangle$ $(s = 0, \pm 1)$. Hyperfine coupling tensors between the central NV center and the nuclear spin are determined from first principles density function theory (DFT) calculations as specified in Refs.~\citep{IvadyPRb2020,Szasz13}.  Spin-relaxation in the excited state is neglected in this study.  Typically 32 cycles are considered, which corresponds to $\approx 0.1$~ms overall simulations time. We note that simulation of longer pumping is possible, however, beyond $0.1$~ms we experience considerable finite-size effects in our model consisting of 127 nuclear spins, find more details in the appendix.  Based on the convergence tests summarized in the appendix, we set the order parameter to 2, meaning that $^{13}$C-$^{13}$C coupling is included in the model between pairs of close nuclear spins, and considered an ensemble of 100 random spin configurations in all cases when $^{13}$C nuclear spin bath is considered. 

For point-defect spin environments we make an assumption that the spin-relaxation time of the spin defects is shorter than the inverse of the coupling strength and the pump rate, thus dynamical polarization of the spin defects due to interaction with the central NV center may be neglected. Omitting optical polarization cycles, we simulate a ground state time evolution of 0.1~ms dwell time to model such systems. For P1 center and NV center spin environments we assume non-polarized and nearly completely polarized states for the spin bath, respectively. Coupling tensors between the central and environmental spin are calculated from the dipole-dipole interaction Hamiltonian. Our ensembles induce 100 random spin defect configurations, each of them consisting of 127 spin defects. Electron spin defects usually possess shorter coherence time than the inverse of the NV coupling strength, therefore, the bath may  be considered uncorrelated and the first-order cluster approximation is appropriate in these cases.\citep{IvadyPRb2020}

\subsection*{Samples and experimental methods}

In our experiments we study  different diamond samples with different defect concentration and $^{13}$C abundance. Table~\ref{tab:sample} summarizes the most relevant properties of all the studied samples. 

\begin{table}
\caption{ Specifications of the samples used in our experiments.}
\begin{ruledtabular}
          \begin{tabular}{cccc} 
Sample      & NV  &  P1 &  $^{13}$C     \\ \hline 
W4     &        $10-20$ ppb         &          1 ppm    & 1.07\%     \\
IS       &         0.9 ppm     &   2 ppm  & 0.03\%    \\
E6      &         2.3 ppm        &       13.8 ppm           &   0.01\% \\
F11    & $<20$ ppm &  $<200$ ppm & 1.07\%  \\
       \end{tabular}
\end{ruledtabular}\label{tab:sample}
\end{table}

We carry out photoluminescence measurements on our samples.  The experimental apparatus includes a custom-built electromagnet which provides magnetic field of 0 to about 110\,mT. The electromagnet can be moved with a computer-controlled 3-D translation stage and a rotation stage. The NV-diamond sensor is placed in the center of the magnetic bore. The diamond can be rotated around the $z$-axis (along the direction of the magnetic field).

\section{Results}
\label{sec:res}

In the following four sections we investigate experimentally and theoretically the most relevant interactions that can have significant effects on the GSLAC PL spectrum.

\subsection{External fields}
\label{sec:extern}

\begin{figure}[h!]
\includegraphics[width=0.9\columnwidth]{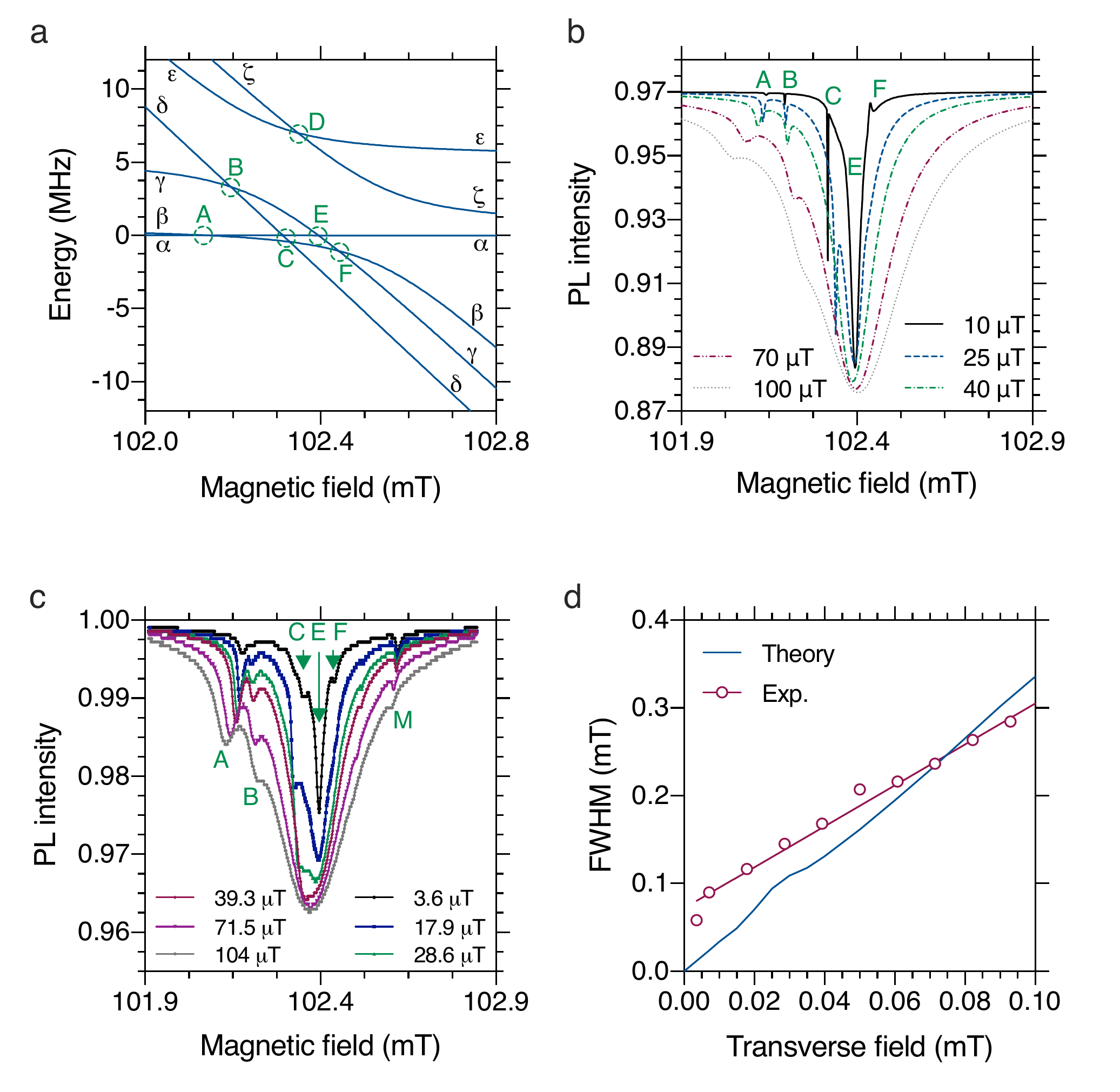}
	\caption{  a) Energy level diagram of the $^{14}$NV center. Greek letters denote the spin eigenenergies and green dashed circles with capital letters denote crossings where external perturbations may open gaps. b) Theoretical PL signal  vs.\  longitudinal magnetic field at the GSLAC  with different values of transverse magnetic field. c) Experimental PL signal obtained from sample IS vs.\  longitudinal magnetic field at the GSLAC  with different values of transverse magnetic field. d) Transverse magnetic field dependence of the full width at half maximum of the central dip E. Maroon open circles, maroon solid line, and blue solid line depict the experimental points, a linear fit, and the theoretical results, respectively. }
	\label{fig:NV}  
\end{figure}

At the GSLAC region the parallel magnetic field is set so that $g_{\text{e}} \beta S_z B_z \approx D$. Due to the large splitting, $ 2D \approx 5.8$~GHz , between $m_{S} = +1$ level and the other electron spin levels and also the dominant spin polarization in the $m_{S} = 0$ spin state, the $m_{S} = +1$ level can be neglected. The relevant energy levels in the vicinity of the GSLAC are depicted in Fig.~\ref{fig:NV}~(a). The corresponding wavefunctions, expressed in the $ \left| S_z, I_z \right\rangle$ basis, are provided in Table~\ref{tab:states}. Besides the hyperfine interaction induced avoided crossings between levels $\gamma$ and $\varepsilon$ and $\beta$ and $\zeta$, one can identify seven crossings. 

In the absence of external field and other spin defects in the environment, the GSLAC PL signal is a straight line with no fine structures at the GSLAC, as the $^{14}$N hypefine interaction does not allow further mixing of the highly polarized $\alpha$ state. External fields, however, give rise to additional spin flip-flop processes that open gaps at the crossings, mix the bright $m_{S} = 0$ and the dark $m_{S} = -1$ spin state and thus imply fine structures in the GSLAC PL spectrum. In Table~\ref{tab:crossings} we list spin flip-flop processes that may take place at crossing A-F, also labeled by two Greek letters that refer to the crossing states. We note that except for the $\alpha\delta$ crossing, all crossings allow additional spin mixing. Precession of the electron and $^{14}$N nuclear spin may be induced by external transverse field. Cross relaxation between the electron and nuclear spin can happen when the initial, near unity polarization of the $\alpha$ state is reduced by external perturbations.

\begin{table}[h!]
\caption{ Energy eigenstates of the $^{14}$NV center at the GSLAC as expressed in the basis of $\left| S_z, I_z \right\rangle$, where $z$ is parallel to the N-V axis. $a$, $b$, $c$, and $d$ are coefficients. }
          \begin{tabular}{cc} \hline \hline 
Label         & Eigenstate      \\ \hline 
$\alpha$    &   $\left| 0, +1 \right\rangle $   \\
$\beta $     &      $ a \left| 0, -1 \right\rangle + b   \left|  -1, 0 \right\rangle$    \\  
$\gamma $     &     $ c \left| 0, 0 \right\rangle + d  \left|  -1, +1 \right\rangle$     \\ 
$\delta$    &   $\left| -1, -1   \right\rangle $   \\
$\varepsilon $     &    $ c \left|  -1, +1 \right\rangle  - d \left| 0, 0 \right\rangle  $        \\ 
$\zeta $     &     $ a \left| -1, 0 \right\rangle - b  \left|  0, -1 \right\rangle$      \\ \hline \hline 
       \end{tabular}\label{tab:states}
\end{table}

\begin{table}[h!]
\caption{Characteristics of the level crossings at the GSLAC of the $^{14}$NV. $\left( \Delta S, \Delta I  \right) $ specifies spin flip-flop processes that may take place at the crossings. }
        \begin{tabular}{cc|c|c} \hline \hline 
 \multicolumn{2}{c|}{\multirow{2}{*}{Crossing}}     &  \multicolumn{2}{c}{ Spin state transitions} \\ \cline{3-4}
& & $\left( \Delta S, \Delta I  \right) $  &  remark \\ \hline 

A & $ \alpha \beta  $     &  $  \left( \pm 1, \mp 1  \right)  $  & cross relaxation \\
B & $\gamma \delta $     &     $\left( \pm 1, \pm 1  \right) $ & cross relaxation \\
C & $\alpha \delta $     &   & no spin state transition allowed  \\
C$^{\prime}$ & $\beta \delta $     &     $\left( \pm 1, 0  \right) $ \& $\left( 0, \pm 1  \right)$  & electron and $^{14}$N spin precession  \\
D & $\epsilon \zeta $     &   $\left( \pm 1, 0  \right) $ \& $\left( 0, \pm 1  \right)$  & electron and $^{14}$N spin precession   \\
 E & $\alpha \gamma $     &    $\left( \pm 1, 0  \right) $ \& $\left( 0, \pm 1  \right)$  & electron and $^{14}$N spin precession  \\
 F & $\beta \gamma $     &    $\left( \pm 1, 0  \right) $ \& $\left( 0, \pm 1  \right)$  & electron and $^{14}$N spin precession    \\   \hline \hline 
       \end{tabular} \label{tab:crossings} 
\end{table}

 As $2 \times 2$ Pauli matrices, $\sigma_x$, $\sigma_y$, and $\sigma_z$, and the $2 \times 2$ identity matrix $\sigma_0$  span the space of $2 \times 2$ matrices, the spin Hamiltonian of any external field acting on the reduced two-dimensional basis of the NV electron spin can be expressed by the linear combination of these matrices at the GSLAC, as
\begin{equation} \label{eq:perturb}
\Delta H =  \delta_x \sigma_x + \delta_y \sigma_y + \delta_z \sigma_z + \delta_0 \sigma_0 \text{.}
\end{equation}
 This means that any time independent external perturbation acting on the electron spin of the NV center at the GSLAC can be described as an effective magnetic field. Therefore, in the following, we restrict our study to transverse magnetic field perturbations that induce spin mixing. This is sufficient to understand the GSLAC PL signal due to external fields.

We study the effects of transverse magnetic field theoretically, in a spin defect-free NV center model, and experimentally, in our 99.97\% $^{12}$C IS diamond sample. In the simulations we evolve the density matrix according to the master equation of the closed system, Eq.~(\ref{eq:master}), over 0.1~ms and calculate the average PL intensity. This procedure allows us to obtain minuscule PL features caused by weak transverse magnetic fields. In Fig.~\ref{fig:NV}~(b) and (c) the theoretical and experimental PL signals are depicted at different transverse magnetic fields. On top of the wide central dip at $B_z = 102.4$~mT, that corresponds to crossing E and to the precession of the electron spin due to the transverse field, altogether four (five) pronounced side dips can be seen on the theoretical (experimental) curves.  The right most dip M in the experiment is related to cross relaxation with the nitrogen spin of other NV centers, for details see section~\ref{sec:res:NV}. Side dips A, B, C, and F are well resolvable in both theory and experiment and we assign them to spin flip-flop processes induced by the transverse magnetic field.  With increasing transverse field, these features shift, broaden, and change amplitude. For example, dips B and C merge with the central dip, while dip A moves away from the central dip. These features are characteristic fingerprints of external transverse fields. We note, that the side dip positions are somewhat different in experiment and theory. We attribute these differences to other, unavoidable couplings in the experiment, e.g.\ parasitic longitudinal and transverse magnetic field, electric fields, and other spin defect.

Transverse-field dependence of the dip position, width, and amplitude can be understood through the energy level structure altered by the transverse magnetic field and the variation of the population of the states induced by additional spin flip-flop processes. As an example we discuss the case of dip C that appears at the crossing $\alpha\delta$ at 102.305~mT. In Table~\ref{tab:crossings} we marked this crossing as not allowed, which is valid in the limit of $B_x \rightarrow 0 $. Indeed, by reducing the strength of the transverse magnetic field the dip vanishes rapidly. At finite transverse field the mixing of the states at the $\beta\delta$ crossing changes the character of level $\delta$ that allows new spin flip-flop processes at $\alpha \delta$. This happens only when the transverse magnetic field is strong enough to induce overlap between the anti-crossing at $\beta\delta$ and crossing at $\alpha \delta$.

Next, we discuss the magnetic field dependence of the linewidth of the central dip E at $B_z = 102.4$~mT, which is relevant for both longitudinal and transverse magnetic-field-sensing applications. We study the full-width at half maximum (FWHM) for the central peak for different transverse magnetic field values both experimentally and theoretically, see Fig.~\ref{fig:NV}~(d). Except the region where  dip C merges with the central dip, the theoretical FWHM depends linearly on the transverse field with a gradient of 3.36 mTmT$^{-1}$. The experimental FWHM depends also approximately linearly on the transverse magnetic field, however, at vanishing transverse magnetic field it exhibits an offset from zero. This is an indication of parasitic transverse fields and other couplings in the experiment. The slope of the experimental curve is measured to be $2.3 \pm 0.1$~mTmT$^{-1}$, which is smaller than the theoretical one. Here, we note that inhomogeneous longitudinal fields can also broaden the peaks at the GSLAC. In this case, the broadening is determined by the variance $\Delta B_{\parallel}$ of the longitudinal field.

\begin{figure}[h!]
\includegraphics[width=0.9\columnwidth]{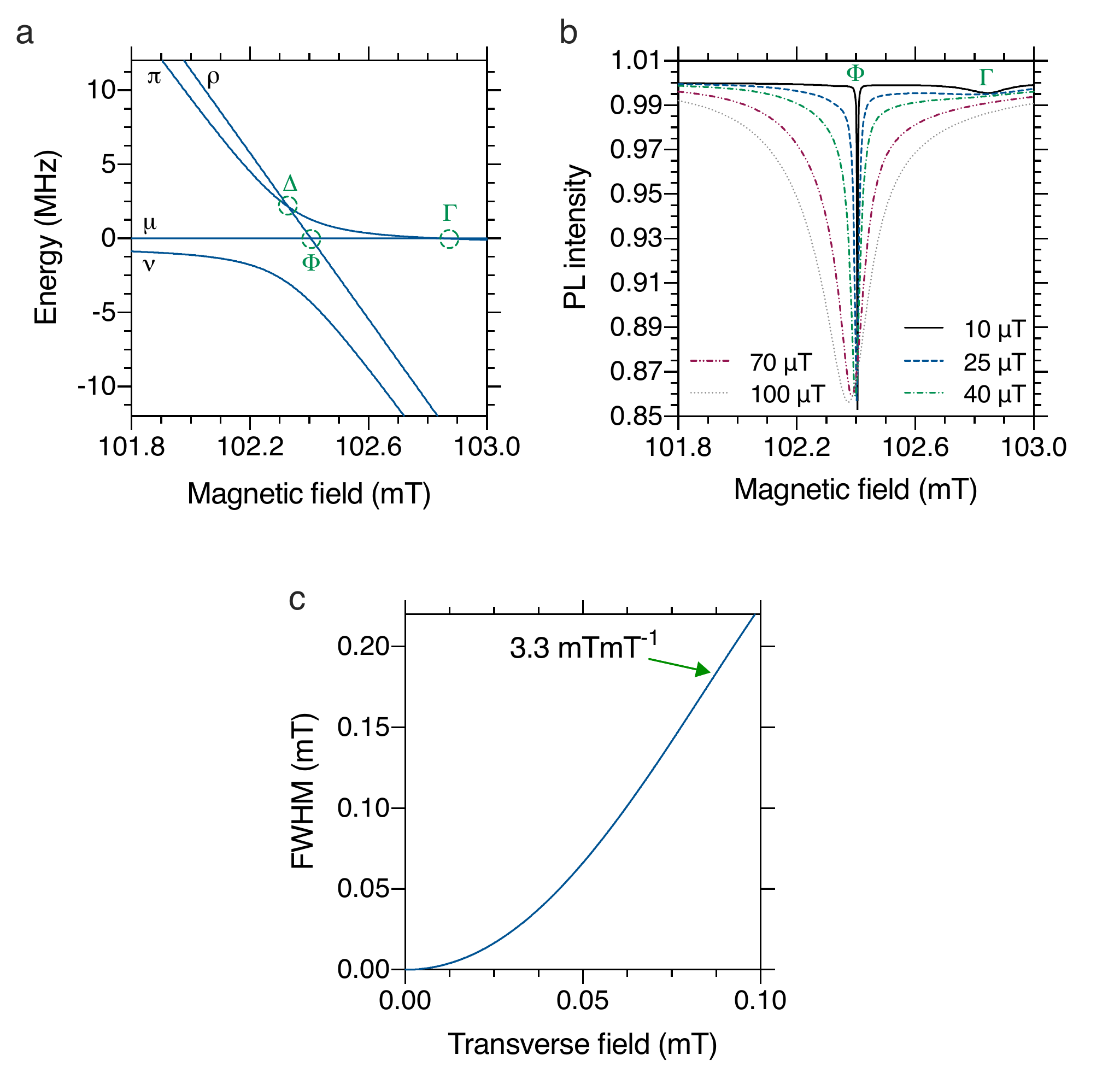}
	\caption{  a) Energy level diagram of the $^{15}$NV center. Greek letters denote the spin eigenenergies and green dashed circles with capital Greek letters denote crossings where external perturbations may open a gap. b) Theoretical PL signal  vs.\  longitudinal magnetic field at the GSLAC  with different values of transverse magnetic field. c) Transverse magnetic field dependence of the FWHM of the pronounced dip at $\approx$102.4~mT. }
	\label{fig:15NV}  
\end{figure}

\begin{table}[h!]
\caption{ Energy eigenstates of $^{15}$NV center at the GSLAC as expressed in the basis of $\left| S_z, I_z \right\rangle$, where $z$ is parallel to the N-V axis. $e$ and $f$ are coefficients. }
          \begin{tabular}{cc} \hline \hline 
Label         & Eigenstate      \\ \hline 
$\mu$    &   $\left| 0, \uparrow \right\rangle $   \\
$\nu $     &      $ e \left| 0, \downarrow \right\rangle + f   \left|  -1, \uparrow \right\rangle$    \\  
$\pi $     &     $ e \left| -1, \uparrow \right\rangle + f  \left|  0, \downarrow \right\rangle$     \\ 
$\varrho$    &   $\left| -1, \downarrow   \right\rangle $     \\ \hline \hline 
       \end{tabular}\label{tab:states-15NV}
\end{table}

Next, we theoretically investigate the PL signal of $^{15}$NV center that is subject to transverse magnetic field of varying strength. The energy level structure of the two spin system is depicted in Fig.~\ref{fig:15NV}~(a), and Table~\ref{tab:states-15NV} provides the energy eigenstates as expressed in the $\left| S_z, {^{15}I_z} \right\rangle$ basis. Besides the hyperfine interaction induced wide avoided crossing of states $\nu$ and $\pi$, three crossing can be seen in  Fig.~\ref{fig:15NV}~(a) that may give rise to PL features in the presence of transverse magnetic field.

\begin{table}[h!]
\caption{Characteristics of the level crossings at the GSLAC of the $^{15}$NV center. $\left( \Delta S, \Delta I  \right) $ specifies spin flip-flop processes that may take place at the crossings. }
        \begin{tabular}{cc|c|c} \hline \hline 
 \multicolumn{2}{c|}{\multirow{2}{*}{Crossing}}     &  \multicolumn{2}{c}{ Spin state transitions} \\ \cline{3-4}
& & $\left( \Delta S, \Delta I  \right) $  &  remark \\ \hline 

$\Delta$ & $ \pi \varrho  $      &     $\left( \pm 1, 0  \right) $ \& $\left( 0, \pm 1  \right)$  & electron and $^{15}$N spin precession  \\

$\Phi$ & $\mu \varrho $    &   & no spin state transition allowed  \\
$\Gamma$ & $\mu \pi $ &     $\left( \pm 1, 0  \right) $ \& $\left( 0, \pm 1  \right)$  & electron and $^{15}$N spin precession    \\   \hline \hline 
       \end{tabular} \label{tab:crossings-15NV} 
\end{table}

Figure~\ref{fig:15NV}~(b) depicts the simulated PL signal of $^{15}$NV center exhibiting two dips on the PL curves. Based on the position of the dips,  the pronounced dip at 102.4~mT can be assigned to the crossing marked by $\Phi$, while the shallow dip at 102.85~mT, observable only at low transverse magnetic fields, is assigned to crossing $\Gamma$. In order to describe the processes activated by the transverse magnetic field at these dips, in Table~\ref{tab:crossings-15NV} we detail the crossings observed at the GSLAC of $^{15}$NV center. As can be seen spin precession is only possible at crossing $\Delta$ and $\Gamma$ and forbidden in first order at crossing $\Phi$. Due to the high degree of polarization in state $\mu$ and the weak spin state mixing at $\Gamma$, spin precession is suppressed to a large degree at crossing $\Delta$ and $\Gamma$. The prominent PL signature  at $\approx 102.4$~mT is enabled by  the interplay of the spin state mixing at crossing $\Delta$ and $\Phi$. For large enough transverse magnetic fields the avoided crossing appears at $\Delta$ overlaps with the crossing at $\Phi$ that enables additional mixing with the highly polarized $\mu$ state. The role of this second order process greatly enhances as the transverse magnetic field increases and eventually gives rise to a  prominent PL dip at the GSLAC.

In Fig.~\ref{fig:15NV}~(c) we depict the transverse magnetic field dependence of the FWHM of the central dip of the GSLAC PL signal of $^{15}$NV center. Due the second-order process involved in the spin mixing, the FWHM curve is hyperbolic like. The derivative of the curve is approaching zero (3.3~mTmT$^{-1}$) for vanishing (large) transverse magnetic field. Due to additional perturbation and field inhomogeneities, we expect that the linewidth of the central dip saturates at a finite minimal value in experiment, similarly as we have seen for the $^{14}$NV center.

The case of $^{15}$NV center demonstrates that second-order processes enabled by the perturbation of the energy level structure can also play a major role at the GSLAC. Eventually, such processes make $^{15}$NV centers interesting for magnetometry applications.

\subsection{ Interaction with $^{13}$C spin bath}

\begin{figure}[h!]
	\includegraphics[width=0.82\columnwidth]{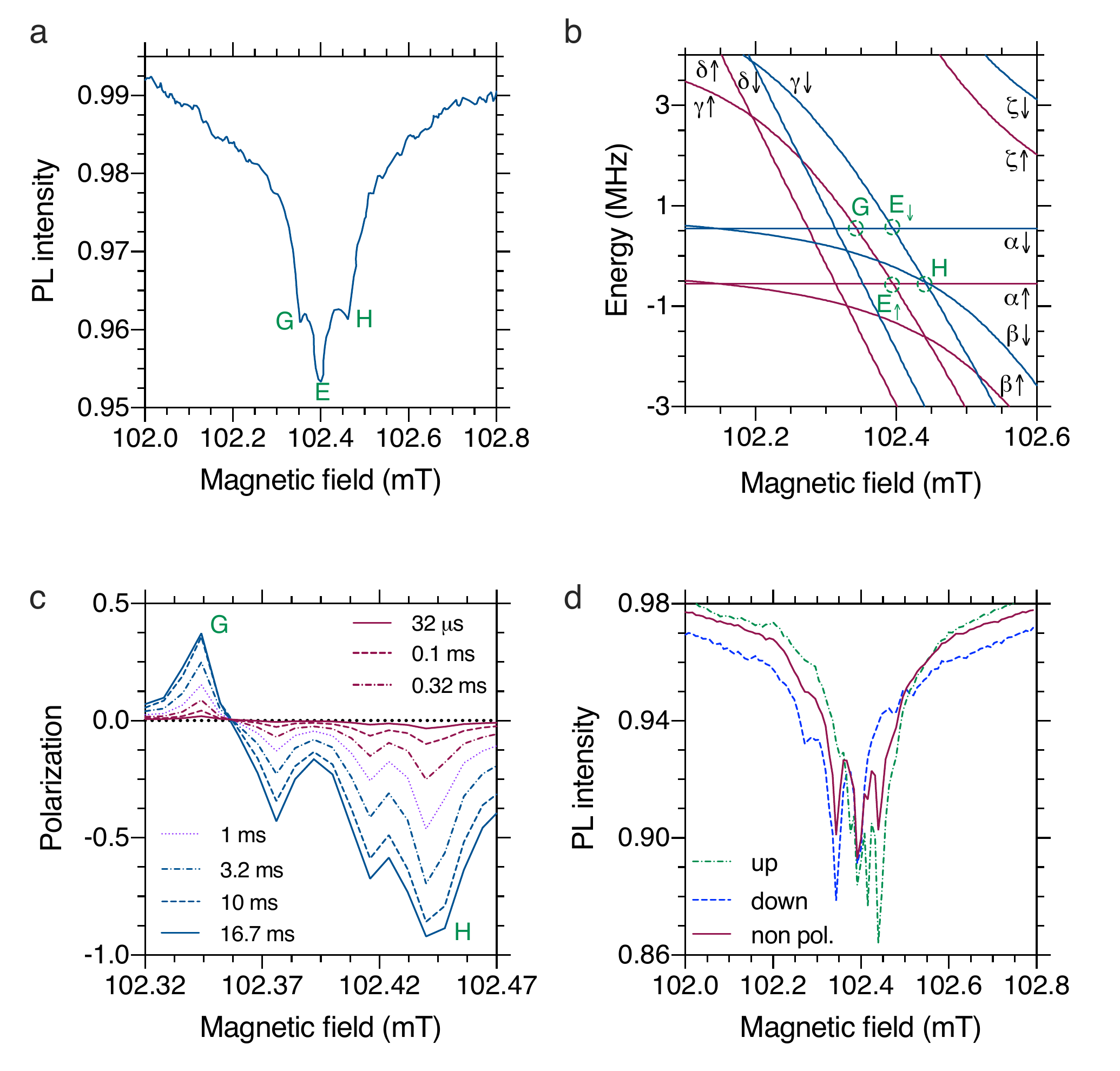}
	\caption{  PL and polarization of $^{14}$NV center-$^{13}$C spin bath system at the GSLAC. a) Measured PL at the GSLAC in sample W4. The PL signal indicates polarization of the $^{13}$C nuclear spin bath, where the sign of the polarization is opposite at the side dips, see text for further discussion. b) Closeup of the energy levels structure of $^{14}$NV-$^{13}$C weakly coupled three spin system at the GSLAC. Greek letter with arrows specify the corresponding states, see Table~\ref{tab:states}, while capital alphabet letters indicate level crossings where hyperfine interaction may give site to additional spin mixing. Maroon and blue curves depict $^{13}$C nuclear spin up and down states, respectively. c) Theoretical ensemble and site averaged polarization of a 128-spin $^{13}$C spin bath obtained after optical pumping of varying duration. d)  Theoretical PL signal obtained by starting from different initial $^{13}$C polarization. The solid maroon curve shows the case of initially non-polarized spin bath, green and blue dashed lines show the PL signal obtained for initially up and down polarized spin bath, respectively. Initial polarization of the spin bath makes side dip amplitudes asymmetric.}
	\label{fig:14NV13C}  
\end{figure}

We study the interaction of $^{14}$NV-$^{13}$C spin bath system at the GSLAC. We record the experimental PL spectrum in our W4 sample of natural $^{13}$C abundance, in which hyperfine interaction with the surrounding nuclear spin bath is the dominant environmental interaction expectedly. A fine structure is observed that exhibits a pair of side dips at $\pm$48~$\mu$T distances from the central dip at 102.4~mT, see Fig.~\ref{fig:14NV13C}~(a).  Similar effects have been recently reported in single-NV-center measurement in 
Ref.~[\onlinecite{BroadwayNatComm2018}].

The phenomenon can be qualitatively understood by looking at the energy-level structure of a $^{14}$NV center interacting with a $^{13}$C nuclear spin at the GSLAC, see Fig.~\ref{fig:14NV13C}~(b), where the level labeling introduced in Fig.~{\ref{fig:NV}} is supplemented with either an up or down arrow depending on the spin state of the $^{13}$C nucleus and the hyperfine interaction is neglected for simplicity. The central dip appears at the place of crossing E$_{\downarrow}$ and E$_{\uparrow}$, where the $\alpha$ and $\gamma $ electron spin states of down and up $^{13}$C nuclear spin projection cross, respectively. Electron spin depolarization and consequent drop of the PL intensity occur at this magnetic field due to the precession of the NV electron spin induced by the effective transverse magnetic field of the nuclear spin. The phenomenon is similar to what we have seen for the case of external transfers fields. The transverse field of the nuclear spin arises from the dipolar hyperfine coupling interaction. According to Eqs.~(\ref{eq:hyp_par2}), the transverse field of individual nuclear spins is proportional to $ \cos \theta \sin \theta$, therefore, it vanishes for $\theta = 0^{\circ}$ and $90^{\circ}$, while it is maximal for $\theta = 45^{\circ}$ and $135^{\circ}$. An important difference between external transverse field and transverse hyperfine field is that the latter varies center-to-center due to distinct local nuclear spin arrangement of individual centers. The varying traverse field induces LAC of varying width at the crossing of $\alpha$ and $\gamma $ levels. Consequently, the central dip observed in an ensemble measurement is a superposition of numerous Lorentzian curves of varying width resulting in a typical line-shape distinguishable from the line-shape observed for homogeneous external fields. 

The left (right) satellite dip corresponds to the crossing G (H) of states $\gamma \uparrow$  and $\alpha \downarrow$  ($\gamma\downarrow$ and $\alpha \uparrow$), where hyperfine term $ S_{\pm} I_{\mp} $ ($ S_{\pm} I_{\pm}  $) may open a gap. According to Eq.~(\ref{eq:hyp_par3}), the strengths of these coupling terms are given by $A_{\parallel} \sin^2 \theta + A_{\perp} \left( \cos^2 \theta + 1 \right)$  and $A_{\parallel} \sin^2 \theta + A_{\perp} \left( \cos^2 \theta - 1 \right)$, respectively. Note that the terms exhibit distinct dependence on the parameters of the hyperfine tensor. Consequently, the left side dip is dominantly due to nuclear spins that are placed on the symmetry axis of the NV center, while the right side dip is dominantly due to nuclear spins that are placed next to the NV center in a plane perpendicular to the NV axis. The PL side dips are caused by mutual spin flip-flops of the electron and nuclear spins that depolarize the electron spin. In turn the nuclear spins can be polarized at the magnetic field values corresponding to the side dips. Due to the different electron and nuclear spin coupling terms efficient at the different side dips, opposite nuclear spin polarization is expected. Indeed, our simulations reveal that the average nuclear polarization $P = \left\langle p_{+1/2} - p_{-1/2} \right\rangle$, where $p_{\chi}$ is the probability of finding individual nuclear spins in state $\left| \chi \right\rangle$, where $\chi = +1/2$ or $-1/2$, and $\left\langle ...\right\rangle$ represent ensemble and bath averaging,  switches as the magnetic field sweeps through the GSLAC, see Fig.~\ref{fig:14NV13C}~(c). These results are in agreement with previous results\citep{HaiJing2013,BroadwayNatComm2018}. 

Dynamic nuclear polarization is demonstrated in Fig.~\ref{fig:14NV13C} (c), where we depict the average nuclear spin polarization obtained after simulating continuous optical pumping of varying duration. The pumping rate is set to 333~kHz in the simulations. It is apparent from the figure that the average nuclear polarization continuously increases as the pumping period extends. The positive and the largest negative polarization dip correspond to the crossing G and H, respectively.  The complicated pattern is however the result of the interplay of different processes that take place at other, not labeled crossings. It is also apparent from the figures that DNP is considerably stronger at the magnetic field corresponding to the right dip.  We also note that in the simulations considerable finite-size effects are observed due to the limited number of spins included in the model, see appendix. Therefore, quantitative results reported in Fig.~\ref{fig:14NV13C} (c) are not representative to the bulk but rather to nano-diamond samples of $\approx 5$~nm size embedding a single, magnetic field aligned NV center. In such small nano-particles nuclear spin diffusion may be negligible, as it is in the simulations.

As the NV center has an effect on the nuclear polarization, the nuclear polarization has also an effect on the NV center, especially on the PL signature at the GSLAC. Similar effects were also seen in single NV center measurements.\citep{BroadwayNatComm2018}
Polarization of the nuclear spins populates and depopulates certain levels that makes the effects of certain level crossing more or less pronounced. In Fig.~\ref{fig:14NV13C}~(d) we model the GSLAC PL spectrum of NV centers interacting with polarized and non-polarized spin baths. Note that the simulation time is set only to 32~$\mu$s in order not to alter the initial polarization significantly. Polarization in nuclear spin up (down) state completely reduces the left (right) dip but in turn enhances the right (left) dip amplitude. Furthermore, additional shallower satellite dips appears. In contrast, the central dip amplitude is affected only marginally by the degree and sign of the nuclear spin polarization. When the spin bath is not polarized initially, i.e. it only polarizes due to optical pumping according to Fig.~\ref{fig:14NV13C}~(c), we observe two side dips of similar amplitudes in the simulations. 

The theoretical PL curve of non-polarized $^{13}$C spin bath in Fig.~\ref{fig:14NV13C}~(d) resembles the experimental curve Fig.~\ref{fig:14NV13C}~(a), however, the amplitude of the side dips is overestimated. As we have seen, this amplitude depends considerably on the polarization of the bath. The relatively small side dip amplitudes in the experiment indicate considerable polarization. We note that the numerical simulation cannot reproduce these curves completely due to finite-size effects observed in the simulations. As mentioned above, DNP at the higher-magnetic-field side dip, that polarizes in the perpendicular plane, is more efficient. Therefore, polarization reaches the side of the simulation box quickly in the simulation, after which the nuclear polarization increases rapidly and reduces the right side dip, that makes the PL side dips asymmetric in amplitude, see appendix.  To circumvent this issue, one may utilize a model including $^{13}$C nuclear spins in a larger, disk shaped volume centered at the NV center.

\begin{figure}[h!]
	\includegraphics[width=0.9\columnwidth]{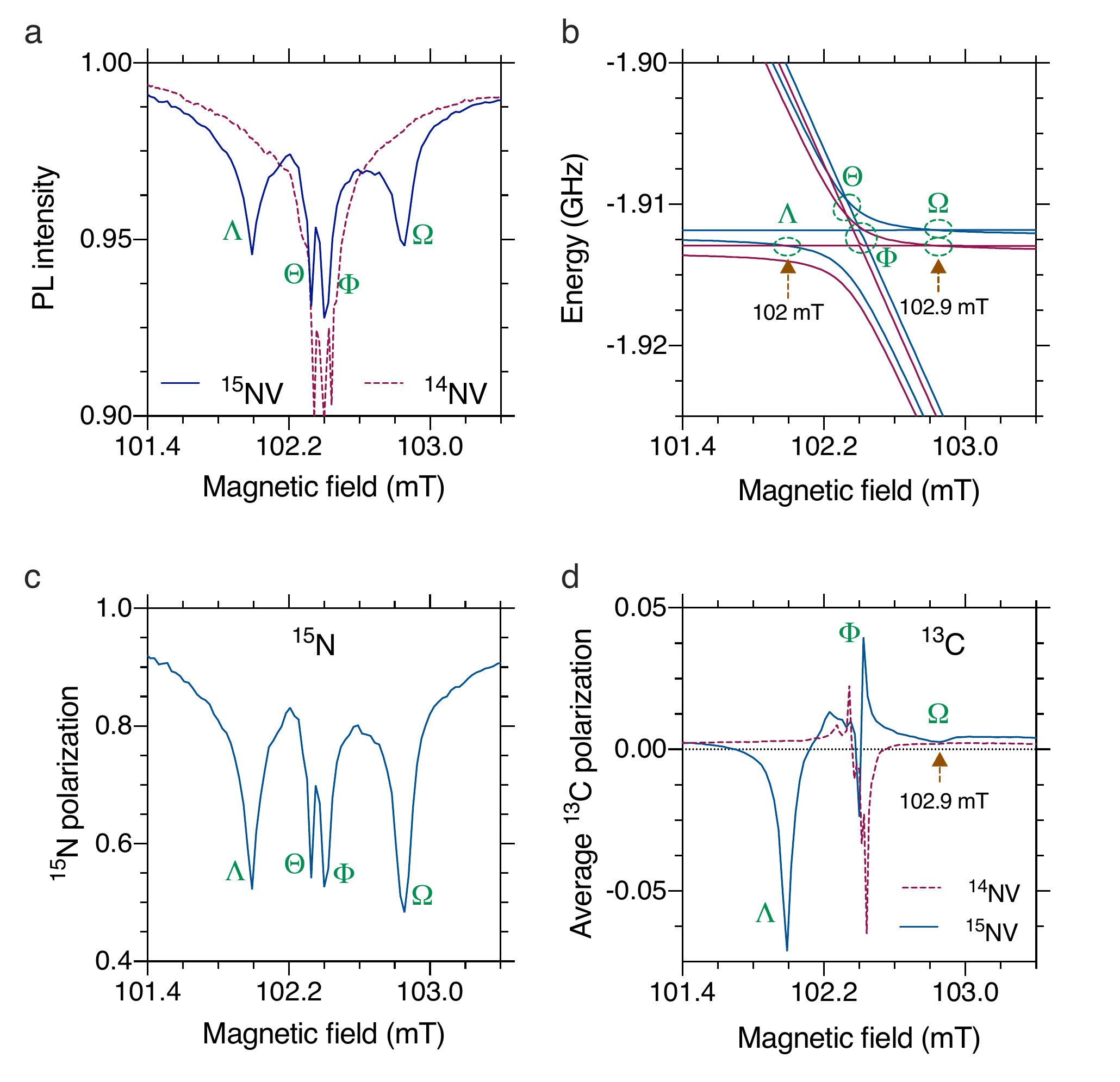}
	\caption{ PL and polarization of $^{15}$NV center-$^{13}$C spin bath system at the GSLAC.   a) PL spectrum of $^{15}$NV-$^{13}$C system (solid blue curve) as compared to the PL spectrum of the  $^{14}$NV-$^{13}$C system (dashed maroon curve).  b) Energy-level structure at the GSLAC. Blue and maroon curves correspond to $\left| -1/2 \right\rangle$ and $\left| +1/2 \right\rangle$ nuclear spin states, respectively. Relevant crossings of the energy levels are indictaed with green dashed circles and labeled by capital letters.  c) and  d) depict the polarization of $^{15}$N nuclear spin and ensemble and site averaged polarization of $^{13}$C nuclear spin bath, respectively. For comparison $^{13}$C nuclear spin bath polarization induced by a $^{14}$NV center at the GSLAC is also depicted in (d). }
	\label{fig:15NV13C}  
\end{figure}

Next, we theoretically investigate the PL spectrum of the $^{15}$NV-$^{13}$C system. The simulated GSLAC PL spectrum, depicted in Fig.~\ref{fig:15NV13C}~(a), reveals a multi-dip fine structure with four distinct dips labeled by capital Greek letters. While the central dip $\Phi$ and its satellite dip $\Theta$ are less prominent compared to the central dip of the $^{14}$NV GSLAC PL signal, we observe two major side dips at larger distances. These dips are similar in amplitude and nearly symmetrical to the central dip, however, their origin is completely different. Figure~\ref{fig:15NV13C}~(b) depicts the energy-level structure of $^{15}$NV-$^{13}$C system and identify the level crossings that are responsible for the observed dips. The central dip and the satellite dip $\Theta$ correspond to the precession of the electron spin driven by the effective transverse field of the $^{13}$C nuclear spin bath. Side dip $\Lambda$ appears at the crossing of states of different $^{13}$C magnetic quantum number, thus hyperfine flip-flop operators may induce mixing between the nuclear and electron spin states. At the place of the dip $\Lambda$, DNP may be realized. Finally, side dip $\Omega$ appears at the crossing of levels of identical $^{13}$C nuclear spin quantum numbers suggesting that here electron spin precession plays a major role.

Figure~\ref{fig:15NV13C}~(c) and (d) depict the polarization of the $^{15}$N nuclear spin and the site and ensemble averaged polarization of the $^{13}$C nuclear spins, respectively. Polarization of $^{15}$N nuclear spin closely follows polarization of the electron spin due to their strong coupling. In absolute terms, the depolarization of the $^{15}$N nuclear spin is twice as large as the electron spin's depolarization indicating that  $^{15}$N nuclear spin plays a role in forming the PL dips. The average polarization of the $^{13}$C spin bath shows distinct signatures, see Fig.~\ref{fig:15NV13C}~(d). Efficient polarization transfer is only possible at the magnetic field corresponding to dip $\Lambda$, where electron spin-nuclear spin mixing is possible. At the central dip $\Phi$ spin coupling gives rise to a sharp alternating polarization pattern. At dip $\Theta$ and $\Omega$ we observe only shallow dips in the $^{13}$C polarization. In order to compare $^{14}$NV and $^{15}$NV DNP processes we depicted in Fig.~\ref{fig:15NV13C}~(d) the averaged nuclear spin polarization obtained for $^{14}$NV center as well. After a fixed 0.3~ms optical pumping, we see that nuclear spin polarization achieved in the two cases is comparable.

As can be seen in Fig.~\ref{fig:15NV13C}~(d) polarization transfer can be as efficient as for the   $^{14}$NV center. The fact that the crossing states at dip $\Lambda$ contain only slight contribution from the $\left| m_{\text{S}} = -1 \right\rangle$ electron spin state suggests that the polarization transfer is suppressed. In contrast, we obtain considerable polarization that we attribute to the absence of competing flip-flop processes at dip $\Lambda$. Processes that could hinder polarization transfer, such as electron spin precession at dip $\Theta$ and $\Phi$, are well separated, in contrast to the case of $^{14}$NV center. These results indicate that besides the most often considered $^{14}$NV center, the  $^{15}$NV center system may also be utilized in MW free DNP applications.

\subsection{Interaction with P1 center and other spin-1/2 point defects}

Diamond often hosts paramagnetic point defects that can interact with the NV centers at the GSLAC.  The spin-1/2 P1 center is a dominant defect in diamond. This defect does not exhibit level crossing at the magnetic field corresponds to the GSLAC, see section~\ref{sec:bg}. Due to the large energy gap between the electron spin states, the central NV center couples non-resonantly to P1 center. This limits the range of interactions to some extent, however, as we show below, efficient coupling is still possible.

\begin{figure}[h!]
	\includegraphics[width=0.9\columnwidth]{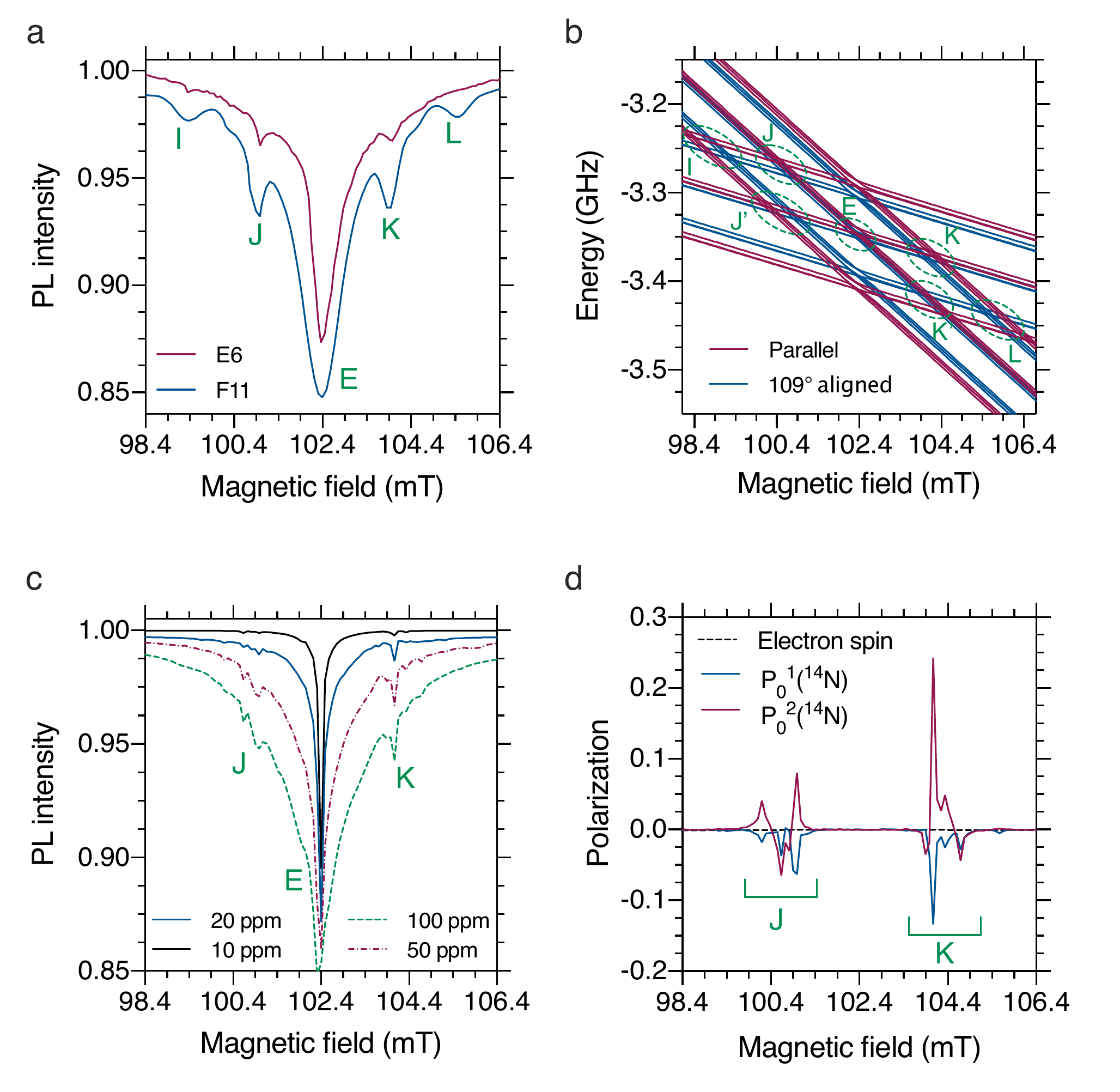}
	\caption{ Interaction with P1 centers at the GSLAC. a) Experimental PL spectrum recorded in two samples of different P1 concentration, see Table~\ref{tab:sample}. PL intensities were scaled to be comparable with each other. b) Energy-level structure of P1 centers. Maroon and blue curves show the case of magnetic field aligned and 109$^{\circ}$ aligned P1 centers, respectively. c) Theoretical PL spectrum for different P1 center concentrations. d) Ensemble and site averaged electron and $^{14}$N nuclear spin polarization of the P1 center.}
	\label{fig:14NV14P1}  
\end{figure}

We study the PL signature of two different samples, E6 and F11, that contain P1 centers in 13.8~ppm and 100-200~ppm concentrations, respectively, see Fig.~\ref{fig:14NV14P1}~(a). Depending on the P1 concentration, we observe either three or five dips in the PL intensity curve. Similar signal has been reported recently in Ref.~[\onlinecite{Armstrong2010,Anishchik2017,Anishchik2019}]. In sample F11 of higher P1 center concentration, two pairs of side dips, I and L and J and K,  can be seen around the central dip E at 102.4~mT. In sample E6 of lower P1 center concentration only side dips J and K can be resolved beside the central dip. Note that the distance of the side dips from the central dip is an order of magnitude larger than in the case of $^{13}$C spin environment, thus these signatures are not related to the nuclear spin bath around the NV center.

In Fig.~\ref{fig:14NV14P1}~(b) we depict the energy-level structure of the magnetic field aligned and 109$^{\circ}$ aligned P1 centers for $m_{\text{P1}} = -1/2$, where one can see three groups of lines for both $m_{\text{S}} = 0 $ and $m_{\text{S}} = -1 $. These separate groups of lines can be assigned to the different quantum numbers of the $^{14}$N nuclear spin of the P1 center. The corresponding states split due to the strong hyperfine interaction.  Note that similar energy level structure can be seen for $m_{\text{P1}} = +1/2$ at 2.8~GHz higher energy. From the comparison of the place of the crossings and the dips in the PL signature we can assign each of the dips to separate crossing regions. The central crossing, labeled E, where the crossing states possess identical P1 center electron and $^{14}$N nuclear spin projection quantum numbers, is responsible for the central dip.  Similarly to the case of the $^{13}$C nuclear spin, the NV electron spin precesses in the effective transverse field of the P1 center caused by the non-secular $S_{\pm} S_{z}^{\text{P1}}$ term of the dipole-dipole interaction. Side dips J and K (I and L) correspond to crossings where the quantum number of the nuclear spin of the P1 center changes by $\pm$1 ($\pm$2). Understanding the mechanisms that give rise to the side dips requires further considerations.  

We theoretically study the PL signal of NV centers interacting with P1 centers ensembles of different concentration, see Fig.~\ref{fig:14NV14P1}~(c). As one can seen, the theoretical curves resemble the experimental ones, however, there are important differences. Even in large P1 concentrations we only see side dips J and K besides dip E in the simulations. The amplitude of the side dips  is also underestimated. Furthermore, the shape of the central peak is different in the simulations and in the experiment, especially in sample E6. The latter can be described by a Lorentzian curve, similarly as we have seen for external fields. This may indicate considerable transverse magnetic field or strain in sample E6.

To understand the mechanism responsible for the side dips, we study the magnetic field dependence of the polarization of the electron spin and the $^{14}$N nuclear spin of the P1 center. The latter can be characterized by $\varrho_0^0$ monopole, $ \varrho_0^1$ dipole, and $\varrho_0^2$ quadrupole moments that correspond to population, orientation, and alignment, respectively. \citep{Auzinsh19}
Orientation and alignment can be obtained from quantities $p_m$ defining the probability of finding the nuclear spin in state $\left| m \right\rangle$ as
\begin{equation}
P_{0}^{1} = \frac{ \varrho_0^1 }{ \varrho_0^0} = \sqrt{ \frac{3}{2} } \frac{p_1 - p_{-1}}{p_1 + p_0 + p_{-1}}
\end{equation} 
and 
\begin{equation}
P_{0}^{2} = \frac{ \varrho_0^2 }{ \varrho_0^0} = \sqrt{ \frac{1}{2} } \frac{p_1 + p_{-1} - 2 p_{0}}{p_1 + p_0 + p_{-1}} \text{,}
\end{equation} 
respectively. The polarization curves as a function of the external magnetic field are depicted in Fig.~\ref{fig:14NV14P1}~(d). Note that the electron spin does not exhibit any polarization. This is due to the fact that the large, 2.8~GHz splitting of the P1 center electron spin states at the GSLAC suppresses flip-flop processes that could polarize the P1 center. The nuclear spin polarization observed in Fig.~\ref{fig:14NV14P1}~(d) might be unexpected, as the electron spin of the P1 center cannot polarize the nuclear spin. Instead, the NV center directly polarizes the nuclear spin of the P1 center. This direct interaction is made possible by the hyperfine coupling that mixes the electron and nuclear spin of the P1 center. Considering only the nuclear spin states, the hyperfine mixing gives rise to an effective $g$-factor that may be significantly enhanced due to the contribution of the electron spin. It is apparent from Fig.~\ref{fig:14NV14P1}~(d) that the nuclear polarization exhibits a fine structure at the magnetic fields that correspond to side dips J and K. This fine structure cannot be resolved in the experimental PL spectrum.

As side dips I and L do not appear in the theoretical simulation we can only provide tentative explanation of these dips. The positions of the dips correspond to magnetic fields where the crossings are related to P1 center nuclear spin state $\left| +1 \right\rangle$ and $\left| -1 \right\rangle$. Therefore, to flip the NV electron spin, the quantum number of the P1 center nuclear spin must change by 2. This may be allowed by the interplay of other spins. For example, $^{13}$C nuclear spin around the NV center or P1 center-P1 center interation may contribute to this process. As side dips I and L are pronounced only at higher P1 center concentrations, we anticipate that the second process is more relevant.

Next, we investigate the linewidth of the central dip E. The varying local environment of the NV centers can induce magnetic field inhomogeneity in an ensemble that broadens the central GSLAC PL dip. This effect may limit the sensitivity of magnetic field sensors. The FWHM for varying P1 concentration $c$, ranging from 10~ppm to 200~ppm, is considered. By fitting a linear curve to the theoretical points we obtain a slope of $ \approx 20$~$\mu$T/ppm. 

Finally, we note that the example of P1 center can be easily generalized to the case of other spin-1/2 point defects. Through the effective transverse magnetic field of the defects, one may expect contribution to the central dip at 102.4~mT. Furthermore, depending on the hyperfine interactions at the paramagnetic defect site, one may observe side dips placed symmetrically beside the central dip. When the point defect includes a paramagnetic isotope of high natural abundance that couples strongly to the electron spin, a pronounced PL signature may be observed at the GSLAC. 

The example of P1 center demonstrates that nuclear spins around spin-1/2 defects can be polarized by the NV center. This phenomena may enable novel DNP applications at the GSLAC. Furthermore, we demonstrated both experimentally and theoretically that GSLAC PL signature depends on the concentration of the spin defect in the vicinity of the NV centers. This realization may motivate the use of GSLAC PL signal in spin defect concentration measurements.

\subsection{NV spin bath}
\label{sec:res:NV}

\begin{figure}[h!]	
\includegraphics[width=0.8\columnwidth]{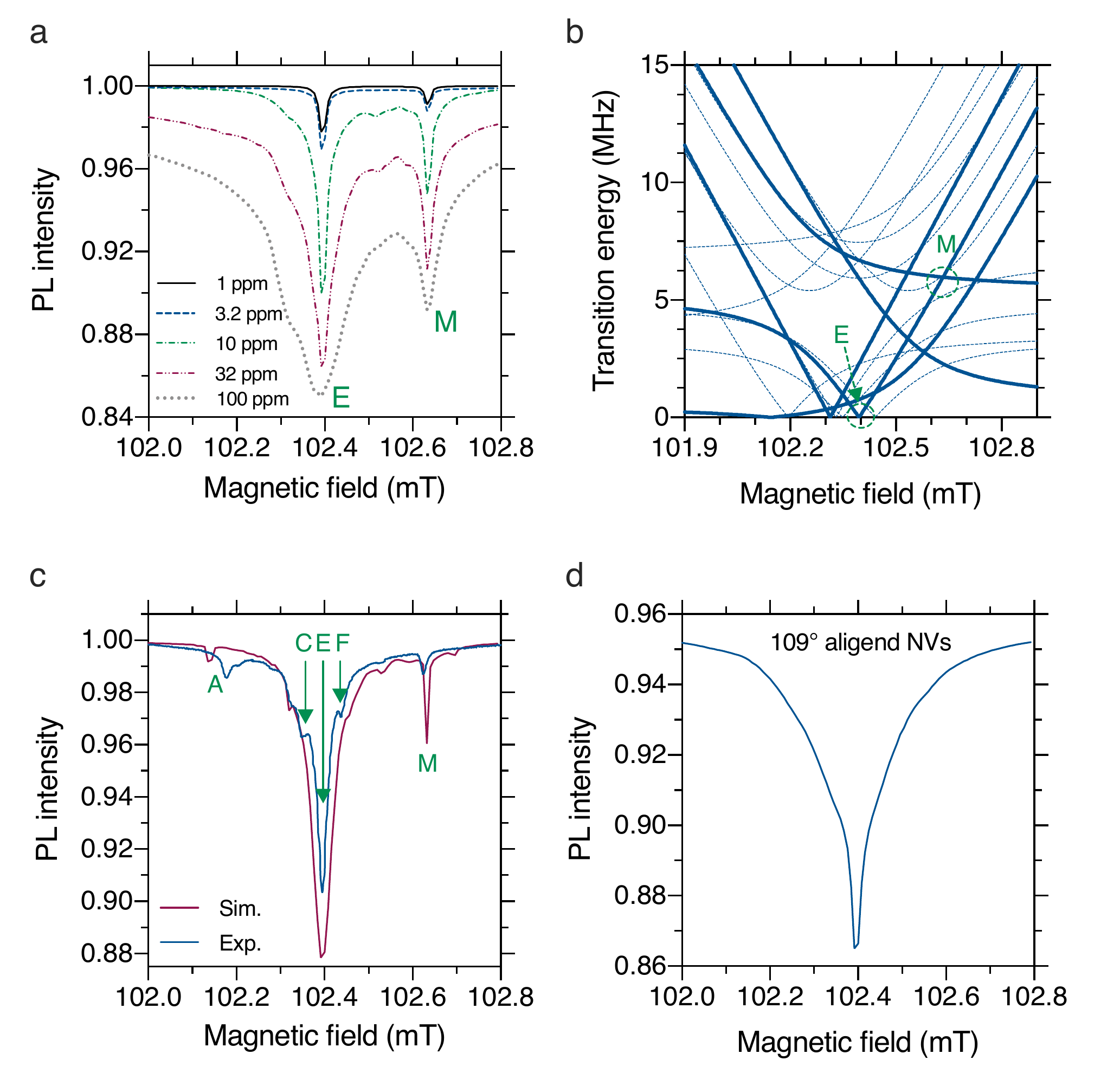}
	\caption{ Photoluminescence of $^{14}$NV center-$^{14}$NV center spin-bath system at the GSLAC.   (a) Simulated PL spectrum of $^{14}$NV center-magnetic field aligned $^{14}$NV spin-bath system at different concentrations ranging from 1~ppm to 100~ppm. (b) Transition energy curves, see text for clarification, at the GSLAC. Wide solid curves highlight transitions associated to the highest populated $\alpha$ state of individual NV centers. Dashed curves show transition energies associated to transitions between less populated energy levels. Green dashed circle labeled by A mark the place of level crossing where the central NV center precesses in the transverse field of other NV centers. Crossing B between the transition energy curves represents a place of effective cross relaxation between the NV centers that gives rise to dip B in the PL scan.  (c) Experimental GSLAC PL spectrum obtained in our IS diamond sample as compared with simulation of 0.5 ppm field oriented NV center spin bath and 15~$\mu$T transverse magnetic field. Signatures of both the transverse field and the parallel NV centers are visible on the experimental and theoretical curves.  (d) Simulated PL spectrum with 109$^{\circ}$ aligned NV center spin bath. }
	\label{fig:14NV14NV}  
\end{figure}

Next, we investigate the case of $^{14}$NV center coupled to a bath of magnetic-field-aligned $^{14}$NV center spins. Note that in this case both the central spin and the bath spins exhibit crossings and anticrossings at the GSLAC. The simulated PL spectrum is depicted in Fig.~\ref{fig:14NV14NV}~(a) for various spin-bath concentrations. As can be seen, two dominant dips, E and M, are observed with additional shoulders appearing at higher concentrations. To track down the origin of the most visible dips, we depict energy curves of allowed spin-state transitions of an individual NV center in Fig.~\ref{fig:14NV14NV}b. Transition-energy curves are differences of energy levels that we label by pairs of Greek letters $\mu\nu$, where $\mu$ and $\nu$ represent levels of individual NV centers, see Fig.~\ref{fig:NV} and Table~\ref{tab:states}. Crossings in the energy level structure of an individual NV center are represented in the transition energy plot by curves approaching zero at the places of the level crossings. Due to the identical level structure of the central and coupled spins, crossings of transition energy curves represent places of efficient cross relaxation between the two NV centers. Cross relaxation can induce depolarization of the central spin that may give rise to dips in the PL spectrum. There are numerous cross relaxation places that can be identified in Fig.~\ref{fig:14NV14NV}~(b). Most of them, however, are not active due to the high polarization of the coupled NV centers. Relevant transition-energy curves associated to the highest populated energy level $\alpha$ in Fig.~\ref{fig:NV} are highlighted in Fig.~\ref{fig:14NV14NV}~(b) by wide solid lines. Dip E appears at the magnetic field where $\alpha\gamma$ transition energy vanishes, ie.\   $\alpha$ and $ \gamma$ states cross, enabling precession of the central NV center in the transverse field of other NV centers, similarly as we have seen for $^{13}$C and P1-center spin baths. We assign the source of dip M in the PL spectrum to a crossing between $\alpha\gamma$ and $\alpha\varepsilon$ transition energy curves.  At the magnetic field of the transition-energy crossing $\alpha\gamma$ transition corresponds dominantly to $\left| 0, +1 \right\rangle \leftrightarrow \left| -1, +1 \right\rangle$ transition, while  $\alpha\varepsilon$ transition corresponds dominantly to $\left| 0, +1 \right\rangle \leftrightarrow \left| 0, 0 \right\rangle$ transition. Therefore, at dip M efficient cross relaxation between the electron and the nuclear spins of the two centers takes place.

To demonstrate $^{14}$NV-$^{14}$NV couplings at the GSLAC experimentally we carry out PL measurement in our IS sample. In Fig.~\ref{fig:14NV14NV}~(c) one can see the experimental PL spectrum obtained at near perfect alignment of the external magnetic field and the NV axis. The residual transverse field is estimated to be $\approx 3.6$ ~$\mu$T. The experimental curve exhibits several dips. To identify each of them we carried out theoretical simulations including additional external transverse magnetic fields of 15~$\mu$T strength. External field is applied solely to the central spin to mimic local transverse inhomogeneities. It is apparent from the comparison that the experimental curve exhibits the signatures of both local field inhomogeneities and interaction with field aligned NV centers in our $^{13}$C depleted sample.

Finally, in Fig.~\ref{fig:14NV14NV}~(d), we theoretically investigate the effect of 109$^{\circ}$ aligned NV center of 10~ppm concentration. These spin defects act like a source of local inhomogeneous transverse field that gives rise to PL signature  similar to the central dip of P1 center induced spin bath. The FWHM of the curve is however twice larger than the FWHM of the P1 center induced PL signature at the same concentration. This is due to the larger magnetic moment of the NV center. 

\section{Discussion}
\label{sec:disc}

The ground-state avoided crossing of the NV center spin states gives rise to a variety  of couplings that imply different behavior of the NV center. We considered the most relevant couplings and demonstrated that each of  them gives rise to a unique PL signature that enables identification of the dominant environmental couplings in a given sample. This may be informative for optimizing defect concentration in samples and experimental setups.

Due to the strong coupling of the NV center to its environment at the GSLAC, an optical signal is produced that makes single and ensemble NV centers interesting for microwave-free sensing and spectroscopy applications. The results collected in this article provide the necessary information for advancing such applications. In spectroscopy, the fine structure and dip positions of the GSLAC PL signal are analyzed. To avoid misinterpretation of the PL signal, it is indispensable to know the signatures of all parasitic interactions that may interfere with the signal to be measured. 

Applications at the GSLAC have focused on the $^{14}$NV center so far. We demonstrate that $^{15}$NV centers can also be utilized in optical applications at the GSLAC with comparable or even superior performance.  $^{15}$NV center based magnetometry, spectroscopy, and hyperpolarization application may be of particular interest. In spectroscopy applications target nuclear spins induced PL dips appear with larger spacing than for $^{14}$NV center, due to the low-angle crossing of the energy levels. This may give rise to better resolution and low sensitivity to other perturbations. 

We demonstrated that, despite the suppressed coupling of the P1 center and NV center electron spins, nuclear spins coupled to a P1 center can be polarized by the NV center at the GSLAC through an effective hyperfine interaction greatly enhanced by the electron spin-electron spin coupling and the hyperfine interaction at the P1 center site. This coupling opens new directions for DNP applications through P1 centers and other spin-1/2 defects at the GSLAC. For example, farther nuclear spin ensembles can be polarized by the NV center without relying on nuclear spin diffusion. This possibility may be particularly important for near surface NV centers that may polarize nuclear spins at the surface though paramagnetic surface defects. In addition, we demonstrated that the  GSLAC PL signal depends considerably on the concentration of paramagnetic point defects, therefore it may serve as a novel tool for measuring spin defect concentration in the vicinity of NV centers. 

Finally, we demonstrated that mutually aligned NV centers can also couple at the GSLAC opening new alternatives for gate operations. While the energy-level structure of coupled NV centers is quite involved at the GSLAC, different spin flip-flop processes resonantly enhance at certain magnetic fields. Depending on the states and the magnetic field, all sorts of operations are possible. We note that $^{15}$NV centers are of great potential in this respect as well. Due to the larger hyperfine splitting and the reduced number of states crossing at the GSLAC, the $^{15}$NV centers may be better controllable.

\section{Summary}
\label{sec:sum}

In summary, we examined, in a joint experimental and theoretical study, most of the relevant interactions at the GSLAC to reveal fine details of the PL signal of NV ensembles. We showed that external fields, $^{13}$C nuclear spin, P1 centers, and other NV centers give rise to unique signatures. These results make identification of the most relevant environmental interactions possible through the GSLAC PL signal. In addition, we provide comprehensive description of all the relevant factors that are needed to be taken into consideration in microwave-free sensing, spectroscopy, and dynamic nuclear polarization applications at the GSLAC.

\section*{Acknowledgments} 

The fruitful discussions with Chong Zu, Konstantin L. Ivanov, and Anton K. Vershovskiy are highly appreciated.  VI acknowledges the support from the MTA Premium Postdoctoral Research Program. VI and IAA acknowledge support from the Knut and Alice Wallenberg Foundation through WBSQD2 project (Grant No.\ 2018.0071). VI and AG acknowledge the Hungarian NKFIH grants No.\ KKP129866 of the National Excellence Program of Quantum-coherent materials project. VI and AG acknowledge support of the NKFIH through the National Quantum Technology Program (Grant No. 2017-1.2.1-NKP-2017-00001). This work was supported by the EU FETOPEN Flagship Project ASTERIQS (action 820394), and the German Federal Ministry of Education and Research (BMBF) within the Quantumtechnologien program (FKZ 13N14439 and FKZ 13N15064), and the  Cluster of Excellence Precision Physics, Fundamental Interactions, and Structure of Matter (PRISMA+ EXC 2118/1) funded by the German Research Foundation (DFG) within the German Excellence Strategy (Project ID 39083149). The calculations were performed on resources provided by the Swedish National Infrastructure for Computing (SNIC 2018/3-625 and SNIC 2019/1-11) at the National Supercomputer Centre (NSC) and by the Wigner RCP.

\appendix

\section{Convergence tests}

\begin{figure}[h!]
\includegraphics[width=0.95\columnwidth]{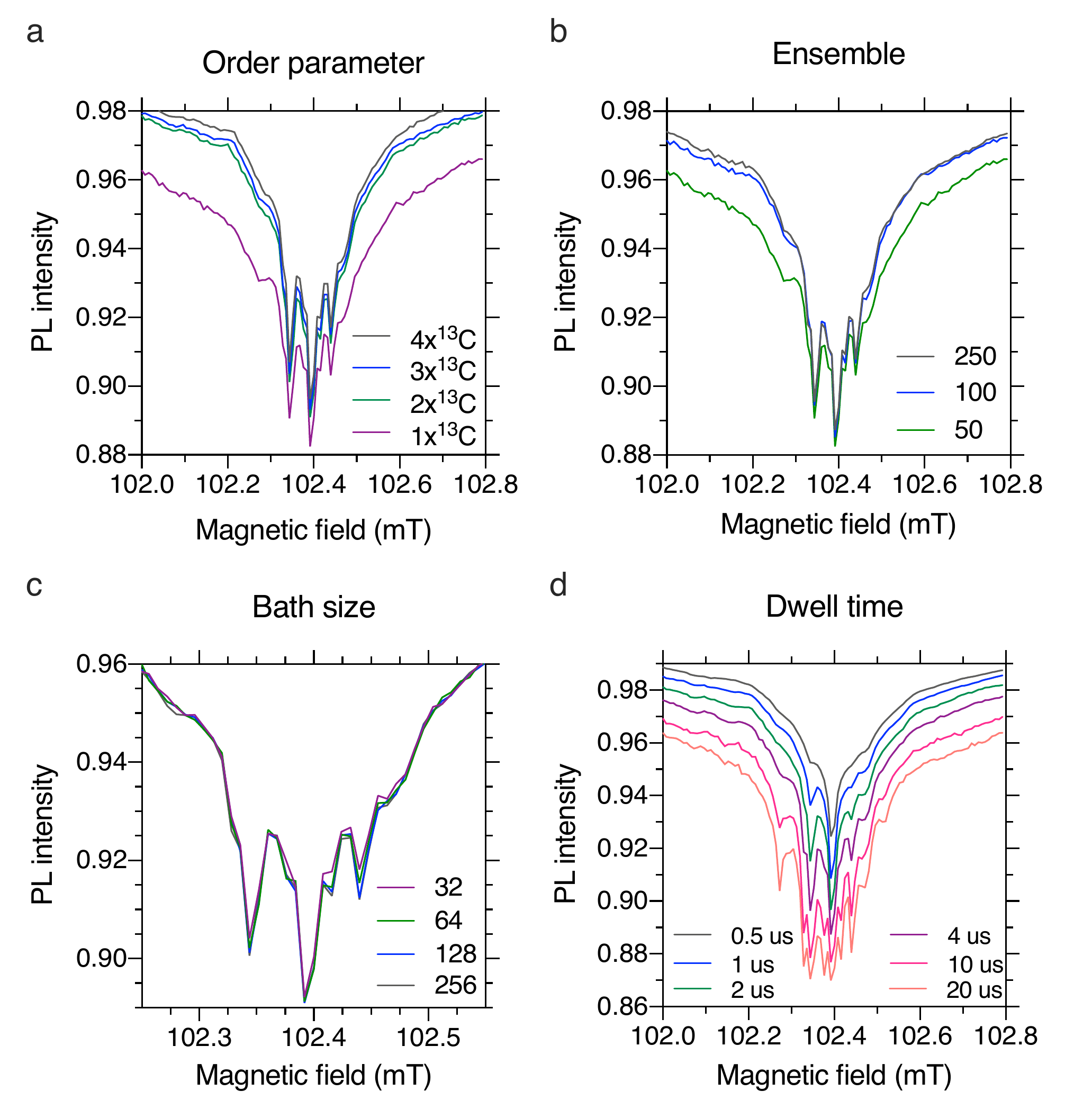}
\caption{ Optimization of the simulation parameters. (a), (b), (c), and (d) show the GSLAC PL signal of $^{14}$NV-$^{13}$C system for different order parameter, ensemble size, bath size, and ground state dwell time, respectively. 
 \label{fig:convtest}  }
\end{figure}

In order to determine the optimal simulation settings, we carry out initial convergence tests. We consider $^{13}$C spin bath, that may couple coherently to the central NV center, due to the long coherence time of the nuclear spins and the relatively strong coupling strength for the closest nuclear spins. 

Neglect of spin-bath correlation effects is the main approximation of the utilized theoretical approach. Spin-bath correlation can be included systematically in the simulations, however, by increasing the order of cluster approximation, i.e. the number of spins included in each subsystem. In Fig.~\ref{fig:convtest}(a), we depict the PL signal obtained for different order parameters, where one can see a significant difference between the case of non-correlated spin bath, $1 \times ^{13}$C, and partially correlated cases, $N \times ^{13}$C, where $N > 1$. Beyond $N = 2$, the PL curves change only slightly, thus we use $N = 2 $ in the simulations of $^{13}$C spin bath. Note that other spin defects considered in the main text include electron spins that usually possess much shorter coherence time, therefore the bath may be considered uncorrelated and  the first-order cluster approximation is appropriate in those cases.

In Figs.~\ref{fig:convtest}(b)-(c), we study ensemble- and bath-size dependence of the GSLAC PL signal. As can be seen 100 and 128 are convergent settings for the ensemble and bath sizes, respectively. Finally, in Fig.~\ref{fig:convtest}(d), we investigate ground-state dwell-time dependence of the PL curves. For increasing dwell time we observe additional fine structures appearing. In the simulations we use 3~$\mu$s dwell time that is a reasonable choice knowing the optical laser power usually used in the experiments.

\section{Pumping-duration dependence of the GSLAC PL signal and  the $^{13}$C  nuclear polarization}

\begin{figure}[h!]
\includegraphics[width=0.95\columnwidth]{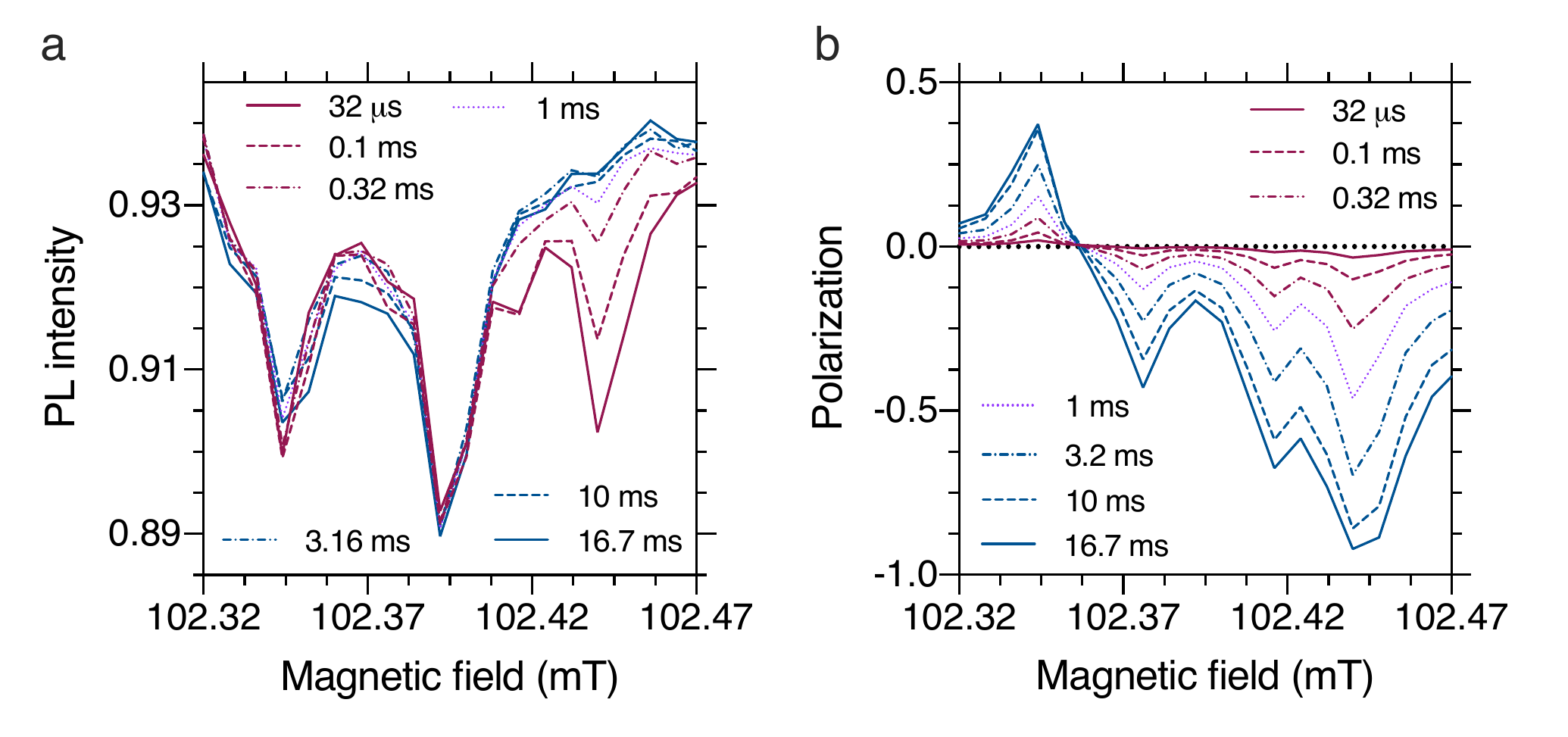}
\caption{  (a) PL signal and (b) $^{13}$C polarization after varying length of optical pumping of a $^{14}$NV center at the GSLAC. 
\label{fig:longrun}  }
\end{figure}

In Figs.~\ref{fig:longrun}(a)-(b) we depict the GSLAC PL signature and the average nuclear polarization of a $^{13}$C bath as obtained after varying length of optical pumping of a $^{14}$NV center. Averaging is carried out over an ensemble of 100 randomly generated NV center-$^{13}$C spin bath configurations, each of which includes 127 $^{13}$C nuclear spins in an arrangement corresponds to natural abundance. With increasing time the right PL side dip reduces rapidly, while the left side dip reduces only moderately in the simulations. The corresponding rightmost and leftmost nuclear polarization peaks in Fig.~\ref{fig:longrun}(b) grows rapidly and modestly, respectively. This shows that the polarization transfer is most efficient at the magnetic field corresponds to the right PL side dip. As the efficiency of the  polarization transfer is varying at the left and right dips, finite-size effects influence the  side dips differently. The different pumping duration dependence of the PL side-dip amplitudes observed in Fig.~\ref{fig:longrun}(a) is attributed to this effect.

\bibliographystyle{apsrev4-1}
%\bibliographystyle{prsty}
%\bibliography{references}

%merlin.mbs apsrev4-1.bst 2010-07-25 4.21a (PWD, AO, DPC) hacked
%Control: key (0)
%Control: author (72) initials jnrlst
%Control: editor formatted (1) identically to author
%Control: production of article title (-1) disabled
%Control: page (0) single
%Control: year (1) truncated
%Control: production of eprint (0) enabled
%

\end{document}